\documentclass[preprint]{revtex4}
\usepackage{graphicx}
\usepackage{bm}
\usepackage{dcolumn}
\usepackage{amsmath}

\def \lleq {\lower0.9ex\hbox{ $\buildrel < \over \sim$} ~}
\def \ggeq {\lower0.9ex\hbox{ $\buildrel > \over \sim$} ~}

\def \beq  {\begin{equation}}
\def \eeq  {\end{equation}}
\def \ber  {\begin{eqnarray}}
\def \eer  {\end{eqnarray}}

\begin{document}
\newcommand{\newc}{\newcommand}

\newc{\be}{\begin{equation}}
\newc{\ee}{\end{equation}}
\newc{\ba}{\begin{eqnarray}}
\newc{\ea}{\end{eqnarray}}
\newc{\bea}{\begin{eqnarray*}}
\newc{\eea}{\end{eqnarray*}}
\newc{\D}{\partial}
\newc{\ie}{{\it i.e.} }
\newc{\eg}{{\it e.g.} }
\newc{\etc}{{\it etc.} }
\newc{\etal}{{\it et al.}}
\newcommand{\nn}{\nonumber}
\newc{\ra}{\rightarrow}
\newc{\lra}{\leftrightarrow}
\newc{\lsim}{\buildrel{<}\over{\sim}}
\newc{\gsim}{\buildrel{>}\over{\sim}}
\title{Probing the dynamical behavior of dark energy}
\author{Rong-Gen Cai}
\email{cairg@itp.ac.cn}
\author{Qiping Su}
\email{sqp@itp.ac.cn}
\author{Hong-Bo Zhang}
\email{hbzhang@itp.ac.cn}
\address{
Key Laboratory of Frontiers in Theoretical Physics, Institute of
Theoretical Physics, Chinese Academy of Sciences, P.O. Box 2735,
Beijing 100190, China}


\begin{abstract}
We investigate dynamical behavior of the equation of state of dark
energy $w_{de}$ by employing the linear-spline method in the region
of low redshifts from observational data (SnIa, BAO, CMB and 12
$H(z)$ data). The redshift is binned and $w_{de}$ is approximated by
a linear expansion of redshift in each bin. We leave the divided
points of redshift bins as free parameters of the model, the
best-fitted values of divided points will represent the turning
positions of $w_{de}$ where $w_{de}$ changes its evolving direction
significantly (if there exist such turnings in our considered
region). These turning points are natural divided points of redshift
bins, and $w_{de}$ between two nearby divided points can be well
approximated by a linear expansion of redshift. We  find two turning
points of $w_{de}$ in $z\in(0,1.8)$ and one turning point in $z\in
(0,0.9)$, and $w_{de}(z)$ could be oscillating around $w=-1$.
Moreover, we find that there is a $2\sigma$ deviation of $w_{de}$
from $-1$ around $z=0.9$ in both correlated and uncorrelated
estimates.

\end{abstract}

\maketitle
\section{Introduction}
It has been more than ten years since our universe was found to be
in accelerating expansion~\cite{Riess:1998cb}. A dominated and
uniformly distributed energy component of the universe, called dark
energy (DE), should be responsible for the acceleration. Many
DE models have been
proposed~\cite{Copeland:2006wr,Caldwell:1997mh,Steinhardt:1999nw,Capozziello:2003tk,Li:2004rb}.
The simplest cosmological model is $\Lambda$CDM model, which
contains a cosmological constant as dark energy.  While $\Lambda$CDM
model is still consistent well with all observational data, a lot of
 efforts have been made to find out whether DE is time-evolving
or is just the cosmological constant. To do that, several
parameterizations of equation of state (EoS) of DE have been
proposed to fit with observational data, such as  the ansatz
$w_{de}=w_0 +w'z$ \cite{Huterer:2000mj}, the EoS expanded by
redshift, and the CPL parametrization
\cite{Chevallier:2000qy,Linder:2002et} $w_{de}=w_0+w_az/(1+z)$,
expanded by scale factor. Both of them contain two free parameters:
$w_0$, the present value of EoS, and $w'$ or $w_a$, represents the
time evolution of EoS. Clearly, constraints of EoS obtained by using
these parameterizations are model-dependent. Given an unreal
assumption of EoS of DE, one may lead to wrong conclusions. Some
model-independent methods have also been
proposed~\cite{Huterer:2002hy,Huterer:2004ch,Hojjati:2009ab,Wang:2009sn},
such as the widely-used uncorrelated bandpower estimates (UBE)
\cite{Huterer:2004ch,Sullivan:2007pd}, in which the redshift is
binned and $w_{de}$ is assumed as a constant in each redshift bin.
Note that the UBE method just approximates the actual $w_{de}$ by an
averaged constant in each bin if DE is dynamical. If there are
sufficient data, $w_{de}(z)$ can be accurately reconstructed.
However, current data could only support a few bins, thus UBE is
always used to test the deviation of $w_{de}$ from the cosmological
constant and used as a supplementary for the parameterizations of
$w_{de}$. Note that the cubic-spline interpolation has also been
 proposed to study the binned
$w_{de}(z)$~\cite{Zhao:2007ew,Serra:2009yp}. However, no convincing
evidence of dynamic DE has been
found~\cite{Zhao:2007ew,Serra:2009yp,Gong:2009ye}. In addition, let
us note that the ansatz, $w_{de}=w_0 +w'z$, of redshift expansion
and CPL parametrization exclude the possibility of an oscillation
EoS, if they are used to fit the whole expansion history of the
universe. While the UBE method needs enough bins to reveal the real
dynamical behavior of DE,  the errors will get larger as the number
of bins increases.

In this paper, we would like to probe the dynamical behavior of
$w_{de}$ by using the linear-spline method. We will approximate
$w_{de}$ in each redshift bin by a linear function $w=w_0+w'z$,
 and require that
$w_{de}(z)$ is continuous in the region under consideration. Since
most of data we used (e.g., SnIa data) are in low redshift, we will
focus on the region of low redshift, such as $z\in(0,0.9)$ and
$z\in(0,1.8)$. In such regions, the width of each redshift bin is small
and the linear expansion could be a better approximation of
$w_{de}(z)$ than a constant in each bin. When fitting with the
observational data, we leave the divided positions of bins $z_i$ as
free parameters. Since the linear function is monotonic, the
best-fitted $z_i$ can represent the turning points of $w_{de}(z)$,
where $w_{de}(z)$ is not linear enough or even non-monotonic (i.e.,
where $d^2w_{de}/dz^2$ departs from zero substantially). Actually we
do find some turning points of $w_{de}$ from observational data, and
the constructed $w_{de}(z)$ just turns its evolution direction at
the best-fitted positions of $z_i$. In this way, we only need to
divide redshift into a few bins, the turning points are natural
divided points of redshift and $w_{de}$ between two nearby points
can be accurately reconstructed by linear expansion. Compared to the
cubic-spline method, the linear-spline (LS) method can find turning
locations of $w_{de}$ more accurately and reduce the errors due to
less the number of bins.  The LS method is also nearly
model-independent, like the piecewise constant and cubic-spline
method. Replacing the linear expansion by CPL parametrization in
each bin, we have reached the almost same results.

For the current status of observational data, LS method may be more
suitable to study $w_{de}$ than the piecewise constant and the cubic
spline method. If DE is dynamical or even oscillating, by using the
LS method it should be more possible to find deviations from the
cosmological constant, at the turning points the deviation from $-1$
should be more explicit. If DE is just the cosmological constant, it
seems more confident if the best-fitted linear expansions construct
an $w=-1$ line, while the oscillation of $w_{de}$ around $w=-1$
could disappear by averaging with the piecewise constant method.
Compared to the piecewise constant case, the only price we pay is
that there is one more parameter in the LS method if the number of bins is the same in two cases.
Compared to the
cubic spline method, the form of $w_{de}(z)$ in each bin only
depends on values of $w_{de}$ at two boundaries, thus the parameters
in $w_{de}(z)$ will not be heavily correlated. Furthermore, the
cubic-spline method seems not suitable for finding the turning
points of $w_{de}$. In all, the LS method could reconstruct $w_{de}$
explicitly by using the least number of bins, and errors of the
parameters from observational data will be small, compared to the
case with more bins.

The paper is organized as follows. In section II we introduce in
detail the method we will use and construct corresponding
cosmological models. In section III, we show how to fit our model
with 397 Constitution SnIa sample \cite{Hicken:2009df}, BAO data from SDSS DR7 \cite{Percival:2009xn},
CMB datapoints ($R, l_a, z_*$) from WMAP5 \cite{Komatsu:2008hk} and 12 Hubble evolution data \cite{Simon:2004tf,Gaztanaga:2008xz}.
The fitting results are presented in section IV. We give our
conclusions and discussions in section V.

\section{Methodology}

To fit models with observational data,  we need to know the form of
Hubble function $H(z)$ (or $E(z)=H(z)/H_0$). In a flat FRW universe
 \be
E^2(z)=\Omega_r^{(0)}(1+z)^4+\Omega_b^{(0)}(1+z)^3+\Omega_{dm}^{(0)}
(1+z)^3+\Omega_{de}^{(0)}F(z),
  \ee where $\Omega_r^{(0)}$,
$\Omega_b^{(0)}$, $\Omega_{dm}^{(0)}$ and $\Omega_{de}^{(0)}$ are
present values of the dimensionless energy density for radiations,
baryons, dark matter and dark energy, respectively, and
$\Omega_r^{(0)}+\Omega_b^{(0)}+\Omega_{dm}^{(0)}+\Omega_{de}^{(0)}=1$.
The energy densities of baryons and dark matter are always written
together as $\Omega_b^{(0)}(1+z)^3+\Omega_{dm}^{(0)}
(1+z)^3=\Omega_{m}^{(0)}(1+z)^3$. The radiation density is the sum
of photons and relativistic neutrinos~\cite{Komatsu:2008hk}:
$$\Omega_r^{(0)}=\Omega_{\gamma}^{(0)}(1+0.2271N_{n}),$$
where $N_{n}$ is the number of neutrino species and
$\Omega_{\gamma}^{(0)}=2.469\times10^{-5}h^{-2}$ for
$T_{cmb}=2.725K$ ($h=H_0/100~Mpc \cdot km\cdot s^{-1}$). The
evolving function $F(z)$ for DE depends on $w_{de}(z)$: \be
F(z)=e^{3\int_0^z\frac{1+w_{de}}{1+x}dx}. \ee For example,
$$F(z)=(1+z)^{3(1+w_0+w_a)}e^{-\frac{3w_az}{1+z}},$$
for the CPL parametrization and $F(z)=1$ for $w_{de}=-1$,
respectively. Here we divide $z\in(0,\infty)$ into $m+1$ bins and
assume $w_{de}(z)$ in the first $m$ bins as \be w_{de}(z_{n-1}<z\leq
z_n)=w_{n-1}+w'_{n}\times(z-z_{n-1})~,~~(1\leq n\leq m) \ee and
require $w_{de}(z)$ to be continuous at divided points: \be
w_n=w_{n-1}+w'_{n}\times(z_n-z_{n-1})~,~~(1\leq n\leq m-1) \ee Note
that here  prime does not represent a derivative, instead $w'_{n}$
is just the slope of the linear expansion in the $n^{th}$ bin. Thus
the independent parameters are \be w_0, w'_1, w'_2, ..., w'_n, ...,
w'_{m}
 \ee
where the total number of parameters is $1+m$ with $m\ge 1$.
Alternatively, we can express  $w_{de}$ as \be w_{de}(z_{n-1}<z\leq
z_n)=w(z_{n-1})+\frac{w(z_n)-w(z_{n-1})}{z_n-z_{n-1}}(z-z_{n-1})~,~~(1\leq
n\leq m) \ee Now the parameters become $w(z_n)$'s, which are values
of $w_{de}$ at the divided points and boundaries $z_n~(0\leq n\leq
m)$ . In this case we have \ba F(z_{n-1}<z\leq
z_n)&=&e^{3\{[w(z_{n-1})-w(0)]+\frac{w(z_n)-w(z_{n-1})}{z_n-z_{n-1}}(z-z_{n-1})\}}
\left(\frac{1+z}{1+z_{n-1}}\right )^{3\frac{w(z_{n-1})(1+z_n)-w(z_n)(1+z_{n-1})}{z_n-z_{n-1}}}\nonumber\\
&&\times(1+z)^3\prod_{i=1}^{n-1}\left(\frac{1+z_i}{1+z_{i-1}}\right
)^{3\frac{w(z_{i-1})(1+z_i)-w(z_i)(1+z_{i-1})}{z_i-z_{i-1}}}
~,~~(1\leq n\leq m) \ea where we have used $z_0=0$. For $w_{de}$ in
the last bin $z\in(z_m,\infty)$, we set it to be a constant $w_L$,
and \be F(z>z_m)=F(z_m)(\frac{1+z}{1+z_m})^{3(1+w_L)} \ee Now the formula for
$H(z)$ is ready.

There is one more thing to be mentioned: once we have fitted our
model with the data introduced in the next section, errors of
$w(z_i)$ are correlated, i.e., the errors of $w(z_i)$ are dependent
on each other. New parameters can be defined by transforming the
covariance matrix of $w(z_i)$, so that errors of new parameters are
decorrelated and  do not entangle with each other. The new
uncorrelated parameters are referred to as the principal
components~\cite{Huterer:2002hy,Hamilton:1999uw}, and they are
directly related to their own locations (unlike the correlated
case).
So errors of the uncorrelated parameters are more interpretable and meaningful.
For more discussions and implications of the uncorrelated
parameters, we refer to the
references~\cite{Huterer:2004ch,Sullivan:2007pd,dePutter:2007kf}. In
section IV, we will show both errors of correlated and uncorrelated
parameters of $w_{de}$. The uncorrelated technique we adopt from
\cite{Huterer:2004ch} is as follows.

1. Get the covariance matrix \be C=\langle WW^T\rangle- \langle W
\rangle \langle W^T\rangle \ee where $W$ is the vector of $w(z_i)$.
The Fisher matrix F is defined by $F=C^{-1}$.

2. Diagonalize the Fisher matrix by an orthogonal matrix O
 \be
F=O^T\Lambda O, \ee
 where $\Lambda$ is diagonal.

3. Define a new matrix $U$ as \be U=O^T\Lambda^{1/2}O, \ee and
normalize $U$ so that the sum of its each row is equal to $1$.

4. Define new parameters $q_i$ by $q=UW$, where $q_i$ are
components of the vector $q$.

Clearly for the case of $w(z)=-1$ (i.e., the cosmological constant
case), we will have $q_i=-1$.
The covariance
of new parameters is
\be \langle (q_i-\langle q_i
\rangle )(q_j-\langle q_j\rangle)\rangle
=\frac{\delta_{ij}}{\sum_a(F^{1/2})_{ia}\sum_b(F^{1/2})_{jb}}.
 \ee
In this way the errors of the new parameters $q_i$ become
uncorrelated.

The uncorrelated parameters $q_i$ are linear combinations of $w(z_i)$, and
the coefficients are just row elements of $U$. The transformation
matrix $U$ constructed in this method ensures that most of the
coefficients are  positive. So most of coefficients are in $(0,1)$.
In this way, the original correlated parameters are weight-averaged,
which leads to the uncorrelated parameters $q_i$.
As a result, if $w_{de}$ is of the quintom
form, the uncorrelated $w_{de}$ always looks  more consistent with
the cosmological constant than the correlated one.

\section{Sets of Observational data }

We will fit our model by employing some observational data including
SnIa, BAO, CMB and Hubble evolution data. The data for SnIa are the
397 Constitution sample~\cite{Hicken:2009df}. $\chi^2_{sn}$ for SnIa
is obtained by comparing theoretical distance modulus
$\mu_{th}(z)=5\log_{10}[(1+z)\int_0^zdx/E(x)]+\mu_0$ (
$\mu_0=42.384-5\log_{10}h$ ) with
observed $\mu_{ob}$ of supernovae: \be
\chi^2_{sn}=\sum_i^{397}\frac{[\mu_{th}(z_i)-\mu_{ob}(z_i)]^2}{\sigma^2(z_i)}
\ee To reduce the effect of $\mu_o$, we expand $\chi^2_{sn}$ with
respect to $\mu_0$~\cite{Nesseris:2005ur}: \be
\chi^2_{sn}=A+2B\mu_0+C\mu_0^2 \label{expand} \ee where \ba
A &=&\sum_i\frac{[\mu_{th}(z_i;\mu_0=0)-\mu_{ob}(z_i)]^2}{\sigma^2(z_i)}\  ,\nonumber\\
B &=&
\sum_i\frac{\mu_{th}(z_i;\mu_0=0)-\mu_{ob}(z_i)}{\sigma^2(z_i)} , \
C=\sum_i\frac{1}{\sigma^2(z_i)} \ea
 Eq.~(\ref{expand}) has a minimum
as
$$\widetilde{\chi}^2_{sn}=\chi^2_{sn,min}=A-B^2/C$$
which is independent of $\mu_0$.
In fact, it is equivalent to performing an uniform marginalization over $\mu_0$,
the difference between $\widetilde{\chi}^2_{sn}$ and the marginalized $\chi^2_{sn}$ is just a constant  \cite{Nesseris:2005ur}.
We will adopt $\widetilde{\chi}^2_{sn}$ as the chi-square between theoretical model and SnIa data.

We will also use the Baryon Acoustic Oscillations (BAO) data from
SDSS DR7 \cite{Percival:2009xn}, the datapoints are
   \be
\frac{r_s(z_d)}{D_V(0.275)}=0.1390\pm0.0037 \ee and \be
\frac{D_V(0.35)}{D_V(0.2)}=1.736\pm0.065 \ee where $r_s(z_d)$ is
the comoving sound horizon at the baryon drag epoch \cite{Eisenstein:1997ik}, and
 \be
D_V(z)=\left [\left(\int_0^z\frac{dx}{H(x)}\right
)^2\frac{z}{H(z)}\right ]^{1/3}  \ee
  encodes the visual distortion
of a spherical object due to the non Euclidianity of a FRW
space-time.

The CMB datapoints we will use are ($R, l_a, z_*$) from WMAP5
\cite{Komatsu:2008hk}. $z_*$ is the redshift of recombination \cite{Hu:1995en}, $R$
is the scaled distance to recombination
  \be
R=\sqrt{\Omega_m^{(0)}}\int_0^{z_*}\frac{dz}{E(z)},
  \ee and $l_a$
is the angular scale of the sound horizon at recombination
  \be
\l_a=\pi\frac{r(a_*)}{r_s(a_*)},
 \ee where $r(z)=\int_0^zdx/H(x)$
is the comoving distance and $r_s(a_*)$ is the comoving sound
horizon at recombination
  \be
r_s(a_*)=\int_0^{a_*}\frac{c_s(a)}{a^2H(a)}da,~~a_*=\frac{1}{1+z_*}
 \ee
  where the sound
speed $c_s(a)=1/\sqrt{3(1+\overline{R}_ba)}$ and
$\overline{R}_b=3\Omega_b^{(0)}/4\Omega_{\gamma}^{(0)}$ is the
photon-baryon energy density ratio.

The $\chi^2$ of the CMB data is constructed as:
\be
\chi^2_{cmb}=X^TC_M^{-1}X
\ee
where
\ba
{X} &=& \left(\begin{array}{c}
l_a - 302.1 \\
R - 1.71\\
z_* - 1090.04\end{array}
  \right)
  \ea
and the inverse covariance matrix
\begin{eqnarray}
 { C_M^{-1}}=\left(
\begin{array}{ccc}
1.8& 27.968& -1.103 \\
27.968& 5667.577& -92.263 \\
-1.103& -92.263& 2.923
\end{array}
\right)
\end{eqnarray}

The fourth set of observational data is $12$ Hubble evolution data
from~\cite{Simon:2004tf} and \cite{Gaztanaga:2008xz}, its
$\chi^2_H$ is defined as
 \be
\chi^2_H=\sum_{i=1}^{12}\frac{[H(z_i)-H_{ob}(z_i)]^2}{\sigma_i^2}.
\ee
Note that redshifts of these data fall in the region $z\in(0,1.75)$.

In summary,
\be
\chi^2_{total}=\widetilde{\chi}^2_{sn}+\chi^2_{cmb}+\chi^2_{bao}+\chi^2_{H} .
\ee

\section{Fitting Results}

\begin{figure}[b]
  \includegraphics[width=6.0in,height=3.6in]{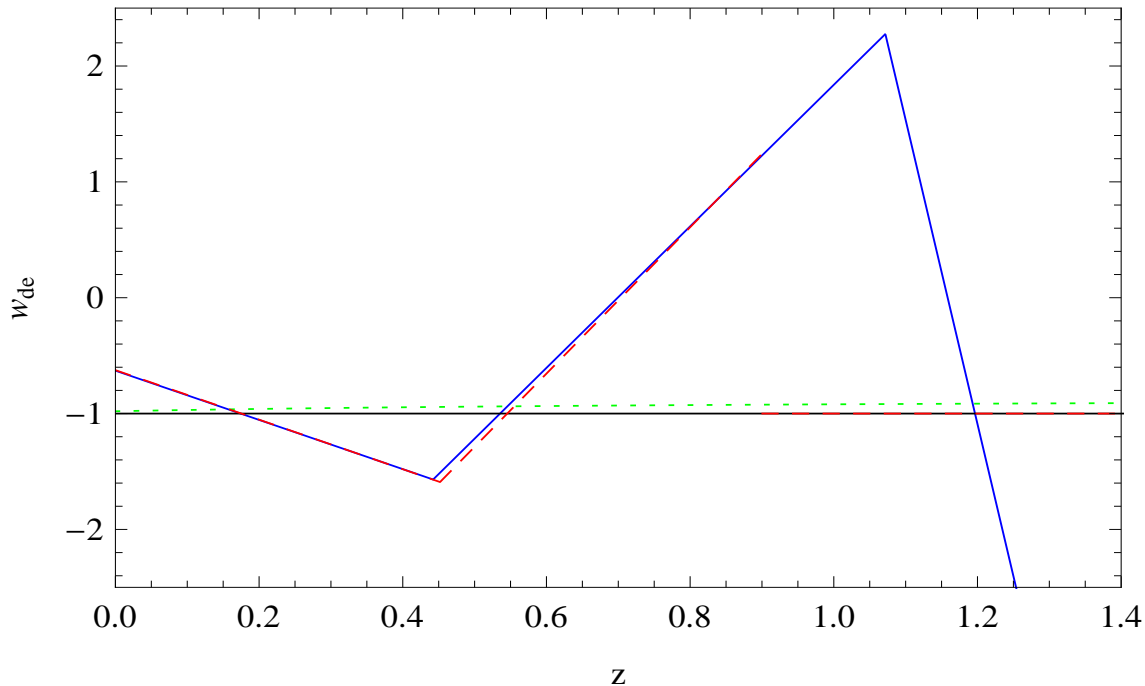}
  \caption{The best-fitted $w_{de}$ for Model I (blue, solid), Model II (red, dashed)
  and the CPL model (green, dotted), the black line is for $w=-1$.} \label{3bin}
\end{figure}

\subsection{Model I}

\begin{table}[b]
\begin{tabular}{c|ccccc}
 \hline
 \hline
Models&~$\widetilde{\chi}^2_{sn,min}$~&~$\chi^2_{cmb,min}$~&~$\chi^2_{bao,min}$~&~$\chi^2_{H,min}$~&~$\chi^2_{total,min}$\\
 \hline
Model I&459.728&0.204&1.494&5.983&467.409\\
CPL&465.621&0.251&1.716&10.818&478.406\\
$\Lambda$CDM&465.759&1.014&1.470&10.850&479.093\\
 \hline
 \hline
\end{tabular}
\caption{The best-fitted $\chi^2$ of four data sets for Model I, CPL model and the $\Lambda$CDM model.}
\end{table}
At first, we divide the whole region of redshift into $4$ bins
(i.e., $m=3$), the divided points and boundaries are $(0, z_1, z_2,
1.8, \infty)$, where $z_1$ and $z_2$ are left as free parameters of
the model, and $0<z_1<z_2<1.8$. In the fourth bin we set $w_L=-1$.
It means that we divide the region with $z\in(0,1.8)$ into $3$ bins
and seek for two possible turning points of $w_{de}(z)$ in this
region. The reconstructed $w_{de}$ of the best-fitted model is shown
in Fig.~\ref{3bin}, which indicates that there exist (at least) two
turning points of $w_{de}$ in $z\in(0,1.8)$ and the best-fitted
values $z_1=0.44$ and $z_2=1.07$. Here $\chi^2_{total,min}=467.409$
for the best-fitted parameters, roughly speaking it is a good
improvement, compared with the corresponding
$\chi^2_{total,min}=478.406$ for the best-fitted CPL model and
$\chi^2_{total,min}=479.093$ for the $\Lambda$CDM model. As shown in
Table I, this improvement of $\chi^2_{total,min}$  is mainly due to
the decrease of $\widetilde{\chi}^2_{sn,min}$ and $\chi^2_{H,min}$. It is not
surprising, as redshifts of the two data sets are distributed in
whole range of $z\in(0,1.8)$, while the BAO data are in the region
$z\leq0.35$ and the CMB data are in the region
$z\in(0,z_*\sim1090)$. The result implies that these two data sets
are quite favor of turnings of $w_{de}$ around $z=0.44$ and $z=1.07$
respectively. This result is consistent with recent UBE of $w_{de}$
\cite{Zhao:2009ti}. While in CPL and $\Lambda$CDM models it is impossible to
have such turnings of $w_{de}$, which leads to the big differences
between $\chi^2_{total,min}$ of Model I and that of $\Lambda$CDM and CPL
models. This result also implies that there exists the possibility
with an oscillating EoS. Note that the error bar of $w_{de}$ in the
third bin is larger than those in the first two bins because there
are much less data points.

We have also divided the region of $z\in(0,1.8)$ into 4 bins, and
found that there is almost no improvement of $\chi_{total,min}^2$ compared
to the case of 3 bins, which indicates there is  no more turning
points of $w_{de}$ in this region.

\begin{table}[h]
\caption{The best-fitted parameters for Model I.}
\begin{tabular}{c|cccccccccc}
 \hline
 \hline
 parameters&~$h$~&~$\Omega_b^{(0)}$~&~$\Omega_m^{(0)}$~&~$z_1$~&~$z_2$~&~$w(0)$~&~$w(z_1)$~&~$w(z_2)$~&~$w(1.8)$~&~ $w_L$\\
 \hline
 best-fitted values&0.688&0.049&0.279&0.44&1.07&-0.63&-1.57&2.28&-16.84&\{-1\}\\
 \hline
 \hline
\end{tabular}
\end{table}
\subsection{Model II}

 As data points with $z>1$ are rather less than those with $z<1$,
  to see clearly the evolution behavior of EoS in the region of low
  redshift, we now focus on the region with $z \in (0,0.9)$, avoiding the possible
  turning point around $z=1$,  and
  set the divided points as: $(0, z_1, 0.9, \infty)$, i.e.,
  \ba w_{de}(z)=\left\{
\begin{array}{cc}
w(0)+\frac{w(z_1)-w(0)}{z_1}z~,& 0<z\leq z_1 \\
  w(z_1)+\frac{w(0.9)-w(z_1)}{0.9-z_1}(z-z_1)~,& z_1<z\leq 0.9 \\
 -1~,& 0.9<z<\infty
 \end{array}\right.
\ea
 In this case, we obtain the best-fitted tuning point
$z_1=0.45$, and the best-fitted $w_{de}(z)$ is shown in
Fig.~\ref{3bin} (the red, dashed line) which almost coincides with
the best-fitted $w_{de}(z)$ of Model I in $z\in(0,0.9)$. This
indicates that the data favor $w_{de}(z)$ to turn its evolution
direction around $z=0.45$, and favor an EoS with crossing the
cosmological constant ($w=-1$)~\cite{Zhang:2009dw}. Then we obtain
$1\sigma$ and $2\sigma$ errors of parameters by
using the MCMC method. Here we have fixed $z_1=0.45$ in the process to obtain
the errors of the parameters.
Note that the errors of the parameters $w_{de}(z_i)$ also represent errors of whole $w_{de}(z)$ in each bin,
the $1\sigma$ and $2\sigma$ errors of $w_{de}(z)$ shown in Fig.~2 are obtained by connecting the corresponding error ranges of $w_{de}(z_i)$.
If another parameter set of $w_{de}(z)$ (as introduced in section II) was used,
one will get the same result as that of Fig.~2.

\begin{figure}[t]
  \includegraphics[width=6.5in,height=3.8in]{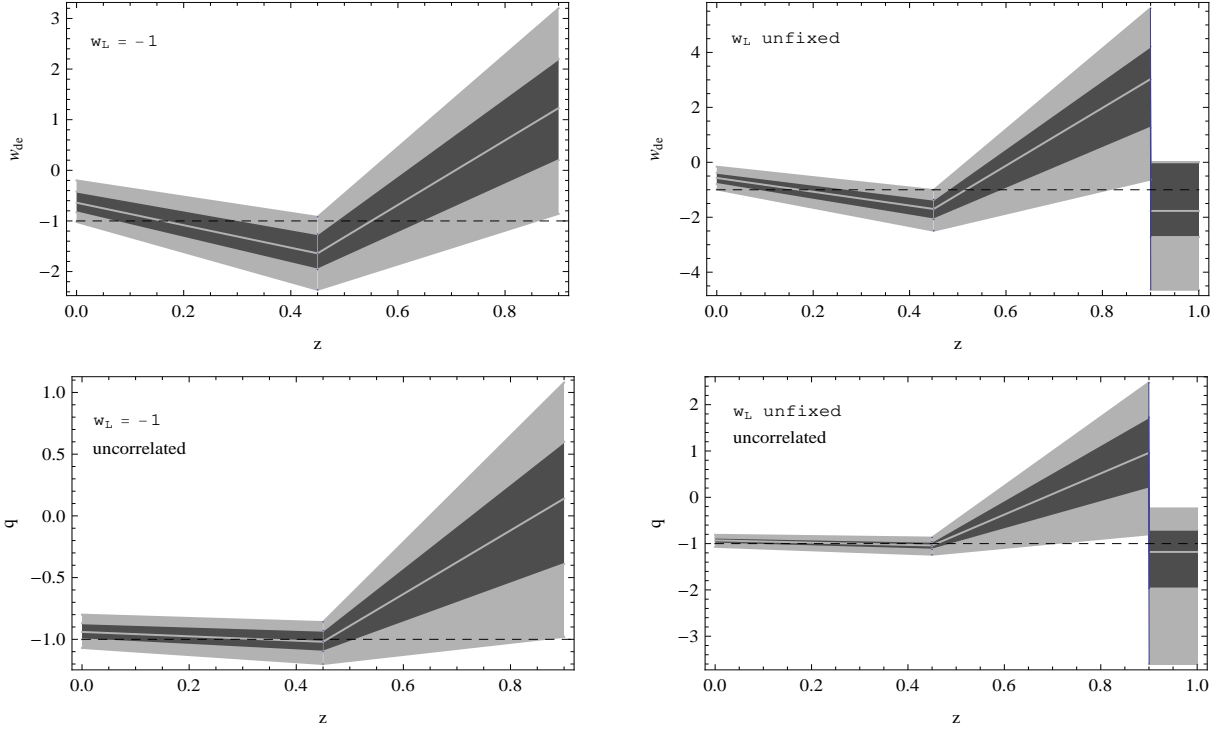}
  \caption{$1\sigma$ and $2\sigma$ errors of $w_{de}$ in Model II.
 Left panels are for the model with $w_L=-1$ and right panels are with $w_L$ floating,
  top panels are for correlated parameters in $w_{de}(z)$ and the bottom panels are for uncorrelated ones.
  }   \label{error}
\end{figure}

We see from the top left panel of Fig.~\ref{error} that there are
deviations of $w_{de}$ from $-1$ around $z=0$ and $z=0.45$ beyond
$1\sigma$, and around $z=0.9$ the deviation is beyond $2\sigma$. We
decorrelate the parameters in $w_{de}(z)$ by using the technique
introduced in section II. The uncorrelated errors are shown in the
bottom left panel of Fig.~\ref{error}. In that case, there are no
more explicit deviations from $-1$ around $z=0$ and $z=0.45$,
however, there is still a $2\sigma$ deviation from $-1$ around
$z=0.9$.

\begin{figure}[t]
  \includegraphics[width=6.5in,height=3.8in]{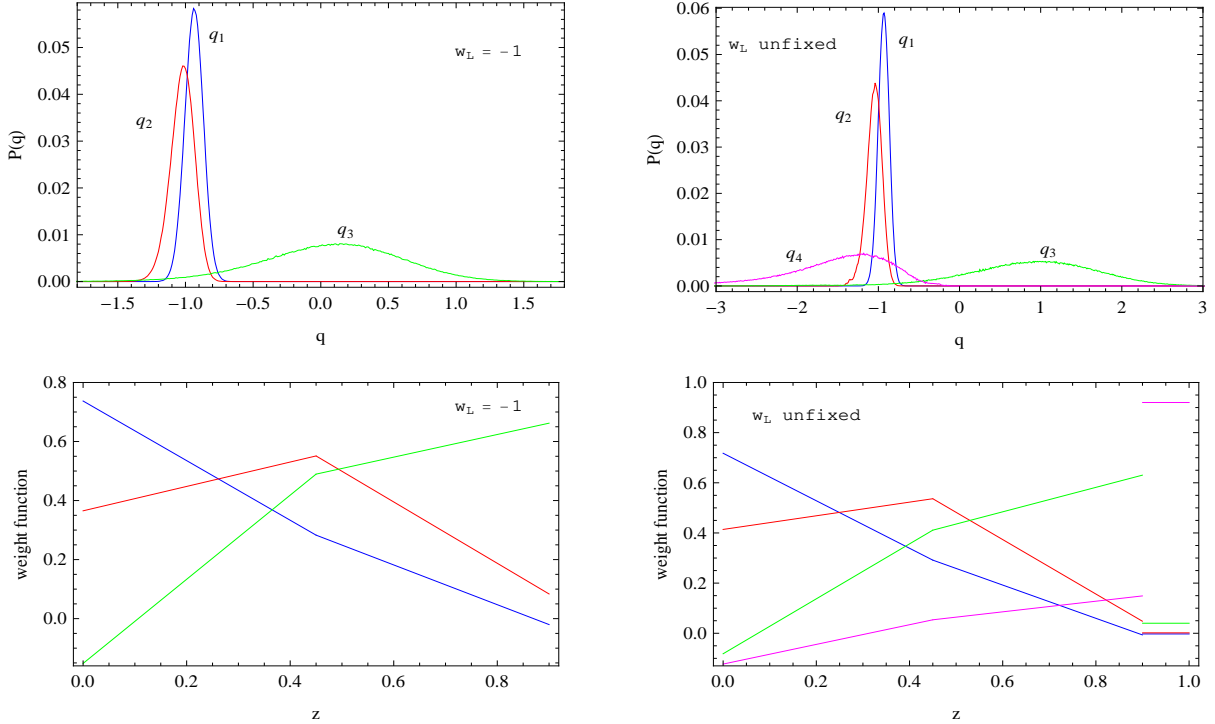}
  \caption{Likelihoods and weight functions of uncorrelated parameters $q_i$ and corresponding weight functions.
  The left panels are for the model with $w_L=-1$ and the right panels are for the case with a floating $w_L$.}
   \label{likelihood}
\end{figure}

One may suspect that the explicit derivations are caused due to the
fact that we have fixed the value of $w_{de}$ as $w_L=-1$ in the
third bin. To check this, we consider the case with a floating $w_L$.
Two right panels of Fig.~\ref{error} show the correlated and
uncorrelated results for the case with the floating $w_L$.  We see
that in this case, there is even larger deviation from $-1$ around
$z=0.9$. We have also used the CPL parametrization to replace the
linear expansion in each bin, and found that the errors are almost
the same as those in the case of the linear expansion and there is
still a deviation of $w_{de}$ from $-1$ around $z=0.9$. We will
extend our discussion of this result in the last section. In Fig.~3
we plot the likelihoods and weight functions of the uncorrelated
parameters $q_i$. The weight functions are constructed from the transformation matrix U,
which show how uncorrelated parameters $q_i$ are
determined. It is shown that the parameters of $w_{de}$ are less correlated than that of
the cubic-spline case~\cite{Serra:2009yp}.

\begin{table}[h]
\caption{The best-fitted parameters and $2\sigma$ errors for Model II with $w_L$ fixed to $-1$ or floating.
ML is for ``Maximum Likelihood'', and the value in \{\} means this parameter has been fixed.}
\small{
\begin{tabular}{c|cccccccc}
 \hline
 \hline
 parameters&~$h$~&~$\Omega_b^{(0)}$~&~$\Omega_m^{(0)}$~&~$z_1$~&~$w(0)$~&~$w(z_1)$~&~$w(0.9)$~&~ $w_L$\\
 \hline
 best-fitted values&0.684&0.050&0.282&0.45&-0.63&-1.59&1.24&\{-1\}\\
 ML and 2$\sigma$ errors&$0.684^{+0.025}_{-0.026}$&$0.050^{+0.004}_{-0.003}$&
 $0.286^{+0.03}_{-0.03}$&\{0.45\}&$-0.64^{+0.44}_{-0.38}$&$-1.64^{+0.72}_{-0.72}$&
 $1.23^{+1.97}_{-2.09}$&\{-1\}\\
  \hline
   best-fitted values&0.687&0.049&0.280&\{0.45\}&-0.59&-1.69&2.54&-1.72\\
   ML and 2$\sigma$ errors&$0.687^{+0.027}_{-0.026}$&$0.049^{+0.004}_{-0.004}$&$0.285^{+0.033}_{-0.030}$&$\{0.45\}$&$-0.57^{+0.41}_{-0.42}$
   &$-1.70^{+0.69}_{-0.79}$&$3.03^{+2.58}_{-3.66}$&$-1.77^{+1.78}_{-2.87}$ \\
 \hline
 \hline
\end{tabular}}
\end{table}

We have also divided $z\in(0,0.9)$ into three bins, to see whether
there exist two turning points of $w_{de}$ in this region. We found
that with the additional $2$ parameters ($z_2$ and $w(z_2)$), there
is almost no improvement of $\chi^2_{min}$, compared to the $2$ bins
case (Model II). This indicates that there is no more turning points
and $w_{de}(z)$ can be well approximated by just two linear
expansions in the region $z\in(0,0.9)$. Of course, there is another
possibility that the current data are not enough to find out more
turning points.

\section{Conclusions and Discussions}

In this paper we have investigated the dynamical behavior of the
EoS of DE in the region of low redshift in a nearly
model-independent way. The redshift in that region is binned  and
$w_{de}$ in each bin is approximated by a linear expansion of
redshift $z$, and in the large redshift region we set $w_{de}$ to
be a constant $w_L$. While fitting the model with some
observational data which include SnIa, BAO, CMB and Hubble
evolution data, we leave the divided points of bins as free
parameters.
If the evolution of $w_{de}$ is not monotonous,  or is not linear
enough in the region under consideration, the best-fitted divided
points will represent the turning points, where $w_{de}$ changes its
evolving direction significantly. In this way we can explicitly reconstruct
$w_{de}$ by using a few bins, and the errors of parameters from
observational data will be small due to the small number of bins.
First we have tried to find the turning points within the region of redshift $z \in
(0,1.8)$, and set $w_L=-1$ in the region $z \in (1.8, \infty)$
(Model I). Our results show that the data favor two
turning points of $w_{de}$ in $z\in(0,1.8)$, and $w_{de}$ may have
an oscillation form \cite{Lazkoz:2005sp}. Our results are consistent with those by the
UBE method in \cite{Zhao:2009ti}.

Since the main data points are in $z\in(0,1)$ and our result in
Model I shows there may be a turning point around $z\sim1$, to see
clearly the dynamical behavior of EoS in that region, we have
focused on the region $z\in(0,0.9)$ in Model II. We have found one
turning point only in $z\in(0,0.9)$, the reconstructed $w_{de}$ in
the best-fitted model is almost the same as that reconstructed in Model I in
$z\in(0,0.9)$.
We have also obtained the errors of
$w_{de}$ at $1\sigma$ and $2\sigma$ in $z\in(0,0.9)$. In both
correlated and uncorrelated estimates with a fixed $w_L=-1$ or a
floating constant $w_L$, we found that there is a $2\sigma$
deviation of $w_{de}$ from $-1$ around $z=0.9$.

It is interesting to see whether the deviation of EoS from $-1$
around $z=0.9$ is physical, or is caused by some unknown technical
causes in fitting. If it is physical, it then clearly shows that DE
is dynamical. But in UBE of $w_{de}$ there seems no such distinct
deviation around $z=0.9$, it may be due to the difference between
the discontinuity of $w_{de}$ in the piecewise constant case and the
continuity in LS case \cite{pre}. In \cite{Serra:2009yp}, where the
cubic-spline method is used, there is also no such an explicit
deviation around $z=0.9$, but it is likely due to its set of EoS in
the last bin $w_L=w(1)$: to fit well with the data of $z>1$, $w(1)$
should be much minus, which would suppress the reconstructed
$w_{de}$ around $z=1$. Of course, it is also possible that such a
big deviation around $z=0.9$ is due to the non-smoothness of
$w_{de}$ at the divided points in our LS method. In the LS method,
$w_{de}$ is continuous, but not smooth at the divided points, i.e.,
its derivative is not continuous at those points. In fact, $w_{de}$
in LS method can be made smooth at the divided points, such as by
the relation~\be
w_{de}(z)=w_0+\sum_{i=1}^{m}\frac{w_i'-w_{i-1}'}{2}(z+\Delta
\ln{\frac{\cosh(\frac{z-z_{i-1}}{\Delta})}{\cosh(z_{i-1}/\Delta)}})
, \ee where $w_{i}'$ is the slope of linear expansion in the
$i^{th}$ bin ($i\geq1$ and $w'_{0}=0$), and $\Delta$ is related to
the smoothed extent at the divided points. With this
parameterization, one can still find out the turning positions of
$w_{de}$ that are favored by observational data, and perturbations
of DE can be calculated.

Furthermore our results are also dependent on the data set we have
used \cite{Sanchez:2009ka}. For example, although there is still a
$2\sigma$ deviation at $z=0.9$ by using the widely-used data set
SnIa + CMB-shift R + BAO parameter A~\cite{Eisenstein:2005su}, now
the best-fitted turning point in $z\in(0,0.9)$ changes to $z=0.39$.
Whatever, from the observational data we have used, a big deviation
of $w_{de}$ from $-1$ around $z=0.9$ is found. Unlike the deviations
around $z=0$ and $z=0.45$, this deviation around $z=0.9$ does not to
be reduced in the uncorrelated estimate. At least, our result shows
that if DE is dynamical it is more possible to find the deviation of
$w_{de}$ from $-1$ around this redshift value.

If the EoS of DE is indeed of an oscillating behavior around $-1$,
it is then not surprising that the cosmological constant always
fits well with observational data because the oscillating
behavior could be smeared in the luminosity distance. However, if
the oscillation region of $w_{de}$ is wide enough (like the case
of our best-fitted $w_{de}$ in Model I), DE may be distinguished confidently
from the cosmological constant by more precise astronomical
observations in the next generation. In addition, let us mention that an oscillating
behavior of $w_{de}$ is also possibly due to some systematic errors in
observational data, or due to some interactions between DE and
dark matter~\cite{He:2009pd}.

\section*{ Acknowledgments}
RGC thanks Y.G. Gong and B. Wang for some relevant discussions.
This work was partially supported by NNSF
of China (No. 10821504 and No. 10975168) and the National Basic
Research Program of China under grant 2010CB833000.

\end{document}